\documentstyle[draft,epsf]{mn}

\def\etal{{\rm et al.\ }}

\def\cf{{\rm c.f.\ }}
\def\gsim{\mathrel{\raise0.35ex\hbox{$\scriptstyle >$}\kern-0.6em 
\lower0.40ex\hbox{{$\scriptstyle \sim$}}}}
\def\lsim{\mathrel{\raise0.35ex\hbox{$\scriptstyle <$}\kern-0.6em 
\lower0.40ex\hbox{{$\scriptstyle \sim$}}}}

\title{Secondary Episode of Star Formation in Elliptical Galaxies}

\author[Kodama \& Arimoto]
{Tadayuki Kodama$^{1,2}$ and Nobuo Arimoto$^{2,3}$\\
 $^1$ Institute of Astronomy, University of Cambridge, Madingley Road,
      Cambridge CB3 0HA, UK\\
 $^2$ Institute of Astronomy, University of Tokyo, Mitaka, Tokyo 181, Japan\\
 $^3$ Observatoire de Paris, Section de Meudon, DAEC, Meudon Principle
      Cedex, 92195, France\\
 e-mail: kodama@ast.cam.ac.uk
}

\date{Accepted for publication in Monthly Notices of the Royal Astronomical
      Society.}

\begin{document}

\label{firstpage}

\maketitle

\begin{abstract}
We put upper limits of secondary burst of star formation in elliptical
galaxies of the Gonzalez's (1993) sample, most of which locates in small 
groups, based on the colour dispersion around the $U-V$ versus
central velocity dispersion relation, and the equivalent width of H$\beta$
absorption.
There are significant number of H$\beta$ strong galaxies which have
EW(H$\beta$)$>2$~\AA, however they do not always have bluer colours in $U-V$.
To be consistent with small colour dispersion of $U-V$,
the mass fraction of secondary burst to the total mass should be less than
only 10~\% at the maximum within recent 2~Gyr.
This result suggests that even if recent galaxy merging produce some
ellipticals, it should not have been accompanied by an intensive star burst,
and hence it could not involve large gas-rich systems.
A capture of dwarf galaxy is more likely to explain dynamical disturbances
observed in some elliptical galaxies.

The above analysis based on the $U-V$ is not compatible
with the one based on the line indices,
which requires that more than 10~\% of mass is present in a 2~Gyr old star
burst to cover the full range of 
the observed H$\beta$ (de Jong \& Davies 1997).
The discrepancy might be partly explained by the internal extinction localized
at the region where young stars form.
However, considering that the H$\beta$ index might have great uncertainties
both in models and in observational data, we basically rely on $U-V$ 
analysis.
\end{abstract}



\section{Introduction}

Elliptical galaxies in clusters are old. Dispersions of colours around 
colour-magnitude ({\it C-M}) relation of cluster ellipticals are small
at low and intermediate redshifts (Bower, Lucey \& Ellis 1992b; Ellis 
\etal 1997). This strongly suggests that almost all stars in ellipticals formed
at a high redshift, at least $z_f>2$--$3$.
Colour change of `red-envelope' galaxies in high redshift
clusters suggests that these galaxies formed at 
$z_f>2$ and evolved passively since then 
(Arag\'on-Salamanca \etal 1993). A similar implication was
given by Mg$_b$ indices of ellipticals in clusters at 
intermediate redshift (Bender, Ziegler \& Bruzual 1996).
Recent analyses of the {\it C-M} relations of cluster E/S0s at high redshift, 
all of which are morphologically classified by HST imaging, 
also require that ellipticals should form at $z_f>2$
(Kodama 1997; Stanford, Eisenhardt \& Dickinson 1997).
An implication derived from these observational evidence is in full 
agreement with a picture predicted by a galactic wind model for the
formation of elliptical galaxies (Larson 1974; Arimoto \& Yoshii 1987;
Kodama \& Arimoto 1997). Elliptical galaxies should form by dissipative 
collapse of proto-galactic massive cloud, during which stars were
born intensively and the star formation
ceased when the gas in a galaxy was expelled due to the super-novae
driven galactic wind. Galaxies then evolved passively
without experiencing any significant event of star formation.

Nevertheless, there is a growing piece of evidence 
showing that elliptical galaxies, in particular in the low density 
environment, might have experienced secondary formation 
of stars in the recent past (e.g., Schweizer \etal 1990; Schweizer 
\& Seitzer 1992; Rose \etal 1994; Bressan, Chiosi \& 
Tantalo 1996; Barger \etal 1996).
Schweizer \& Seitzer (1992) analysed colours and line indices of E/S0 
galaxies in the field as well as in groups and found that $U-B$ and $B-V$ 
become systematically bluer and H$\beta$ indices become stronger 
as the number of fine structure increases, which suggests that the
secondary episode of star formation is related to dynamical
disturbances of galaxies, probably caused by mergers and/or 
interactions with other galaxies (see also Schweizer \etal 1990). 
Hierarchical clustering models of galaxy formation based on the CDM hypothesis
(Kauffmann, White \& Guiderdoni 1993; Cole \etal 1994) 
suggest that the star formation took place recently
in elliptical galaxies due to galaxy merging. 
The CDM based models predict significant fraction of star burst at
$z_f < 1$ in elliptical galaxies especially in the field.

With a help of population synthesis model by Worthey (1994),
de Jong \& Davies (1996) have recently analysed the H$\beta$ and [MgFe] 
indices of elliptical galaxies, spectra
of which were taken by Gonz\'alez (1993). Most of galaxies are the
members of small groups. 
They estimate the amount of star burst 
that is required to reproduce the H$\beta$ 
and disk-to-total light ratio (D/T). The resulting fraction of 
young stars is larger than 10~\% if they are 2~Gyr old.

Barger \etal (1996) demonstrate that at least 30~\% of cluster
galaxies in AC 103, AC 114, and AC 118 ($z=0.31$), seem to have undergone a
secondary burst of star formation during the last $\sim$ 2~Gyr prior to 
the epoch of observation, though many of them are regular E/S0s.
This implies that the frequency of secondary burst is very 
high even for cluster ellipticals, where the galaxy-galaxy merger 
is supposed to be less frequent than in the field.

Independently, broad band colours can put much stronger constraints on 
the nature of secondary star burst in cluster E/S0s. Bower \etal (1992b) 
analysed $U-V$ colour dispersions around the {\it C-M} relations of Coma and Virgo 
clusters, and found that the amount of mass involved in the secondary burst
should be less than 10~\% if the age of burst
population is 5~Gyr. If a younger age of 2~Gyr is assumed,
the burst population should be less than only a few percent.

The problem arising here is an apparent inconsistency between the two 
types of analyses; the line indices against the colour dispersion. The 
lines always indicate larger fraction of the secondary burst population.
Recent studies tend to heavily rely on the line indices data to
derive the contribution of the secondary burst population
(e.g., Worthey 1994; de Jong \& Davies 1996). However, the burst strength
thus derived should also explain the observed $U-V$,
since at least the synthesized broad band colours are much more reliable
than the line indices which are usually calculated independently
of colours by using empirical polynomial fitting functions
(e.g., Worthey 1994; Vazdekis et al. 1997).
In this paper, therefore, we analyse $U-V$ and
H$\beta$ indices of ellipticals in Gonz\'alez (1993) sample,
and derive new constraints on the amount of young population.
 
In \S 2 we compile the data used in this study. In \S 3 we
describe the method to compute models for elliptical galaxies
with the secondary star burst.
The resulting
constraints for the star formation history in ellipticals are given 
in \S 4. In \S 5 we discuss possible causes for the discrepancy between
$U-V$ and H$\beta$ analysis, and conclude this paper.
 
\section {Data}

\subsection {Source}

We have analysed a whole sample of 41 elliptical galaxies to which
Gonz\'alez (1993) gives measurements of the conspicuous Lick line
strengths. H$\beta$ line strengths (at $\lambda \sim 4860$~\AA) and central
velocity dispersions $\sigma_0$ (within the inner 5 arcsec) are taken from
Gonz\'alez (1993) and broad band colours are taken from RC3
(de Vaucouleurs et al. 1991). 
Following Schweizer \& Seitzer (1992), we define $(U-V)_{e,0}$ as
\begin{equation}
 (U-V)_{e,0} = (U-V)_{e} + [(U-V)_{T,0}-(U-V)_{T}],
\end{equation}
where subscript $T$ refers to global colours, subscript $0$ indicates
colours corrected for the extinction due to interstellar dust in our
Galaxy as well as for the redshift, and 
subscript $e$ refers to average colours within an effective radius.
Internal extinction is neglected because they are not given in RC3 
but an argument is given later on the possible influence of the 
internal extinction.
Observational errors are taken into account in $(U-V)_{e}$, 
but errors involved in the correction procedure in deriving $(U-V)_{e,0}$ are
not included because they are not given in RC3.
Therefore the errors of $(U-V)_{e,0}$ could be substantially 
larger than those given in Figs.~\ref{fig:uv},\ref{fig:uv_tb2},
\ref{fig:uv_hbeta_tb2}, and \ref{fig:uv_av_tb2}.
This would not affect our arguments significantly,
however, since our main concern is to understand
why the dispersion in colours is so small along the $(U-V)_{e,0}$
versus $\log \sigma_0$ relation (see Fig.~\ref{fig:uv})
while some galaxies are showing 
strong evidence of young stars in the measured 
H$\beta$ lines. Errors in the velocity 
dispersion are very small ($\sim$ 1~\%; Gonz\'alez 1993), 
and thus are ignored.
 
\subsection {H$\beta$ strong galaxy}

\begin{figure}
\begin{center}
  \leavevmode
  \epsfxsize 1.0\hsize
  \epsffile{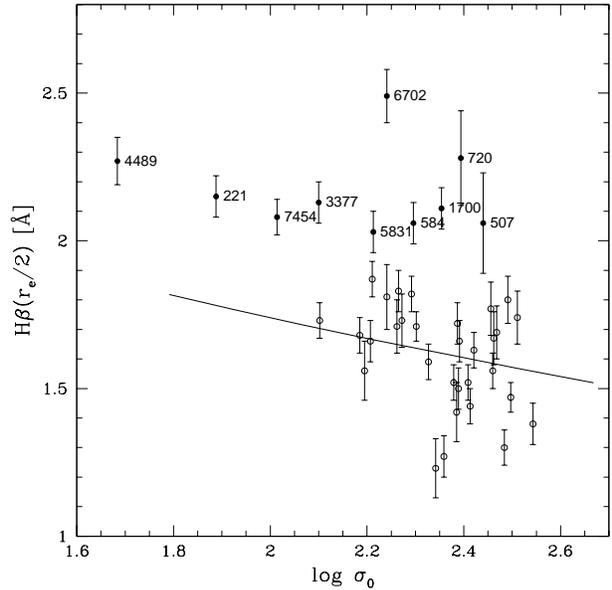}
\end{center}
\caption
{The H$\beta$ versus $\log \sigma_{0}$ diagram for 41 elliptical galaxies.
 Data are taken from Gonz\'alez (1993). The filled circles show the H$\beta$
 strong galaxies to which the NGC number is attached.
 The solid line shows a sequence of galactic wind models (metallicity
 sequence) for elliptical galaxies calculated by Kodama \& Arimoto (1997).}
\label{fig:hbeta}
\end{figure}

The 10 galaxies among Gonz\'alez's sample show strong H$\beta$ absorption,
which may be due to contamination of some young stars in these galaxies.
In analogy to H$\delta$ strong galaxies defined by Couch \&
Sharples (1987), we define H$\beta$ strong galaxies as 
those having equivalent width
EW(H$\beta$) larger than 2~\AA.
Gonz\'alez (1993) gives H$\beta$ indices for three different
apertures; nuclear 5 arcsec, $r_{e}/8$, and $r_{e}/2$,
where $r_e$ is the effective radius of elliptical galaxies.
We adopt the $r_{e}/2$ aperture for H$\beta$ indices, since it is the
closest to $r_{e}$ within which the colours in RC3 are defined.
This would minimize the effect of H$\beta$ line strength gradient 
which is sometimes shown for elliptical galaxies (Gorgas, Efstathiou
\& Arag\'on-Salamanca 1990; Davies, Sadler \& Peletier
1993; Gonz\'alez 1993). Figure~\ref{fig:hbeta}
shows the H$\beta$ indice versus central 
velocity dispersion ($\log \sigma_{0}$) for the Gonz\'alez's
whole sample. The filled circles represent the H$\beta$ strong galaxies. 
The NGC number is attached to each of them. The solid line corresponds to a
sequence of model ellipticals (see \S~3). Note that the
H$\beta$ strong galaxies are away from the sequence more than 0.3~\AA.

\subsection {$U-V$ colour versus velocity dispersion}

\begin{figure}
\begin{center}
  \leavevmode
  \epsfxsize 1.0\hsize
  \epsffile{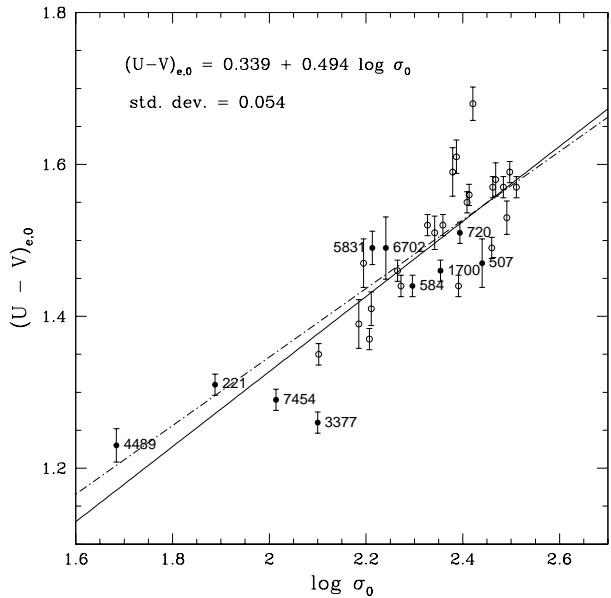}
\end{center}
\caption
{The $(U-V)_{e,0}$ versus $\log \sigma_{0}$ diagram for E/S0 galaxies.
 H$\beta$ strong galaxies are shown by the filled circles to which the NGC
 number is attached. The solid line gives the regression line for the data,
 and the coefficients are indicated on the head of the figure as well as
 the standard deviation of colours. The regression line for Coma and Virgo
 ellipticals observed by Bower \etal (1992b) is also shown by the 
 dot-dashed line.}
\label{fig:uv}
\end{figure}

Figure~\ref{fig:uv} shows the $(U-V)_{e,0}$ versus $\log \sigma_{0}$ 
(km s$^{-1}$) diagram for 33
ellipticals. 8 galaxies among Gonz\'alez's sample are not shown in
Fig.~\ref{fig:uv} because $(U-V)_{e,0}$ are not available in RC3. 
The H$\beta$ strong galaxies are indicated by filled circles. We note 
that a well defined correlation between $(U-V)_{e,0}$ and
$\log \sigma_{0}$ is clearly seen (the {\it C}-$\sigma$ relation)
in this diagram.
The solid line gives a regression line for the sampled galaxies.
The regression line to elliptical galaxies in Coma and Virgo clusters
(Bower \etal 1992b) is also shown by the dot-dashed line. Since the
$U$-band filter used by Bower, Lucey \& Ellis (1992a) is slightly 
different from the standard Johnson system, 
we apply small corrections, although it is negligibly small compared to the
errors in colours ($\Delta(U-V) \simeq 0.01$ mag).
In spite of a difference in the aperture size between the two photometries, 
agreements both in slope and zero-point are remarkable.
The standard deviations are also comparable;
0.046~mag for Bower \etal (1992b) and 0.054~mag for our sample. 
True that most of the Gonz\'alez galaxies locate in small groups,
yet they have similar {\it C}-$\sigma$ relation to that of cluster ellipticals.
Considering the fact that cluster ellipticals have an almost universal
metallicity sequence as a function of galaxy luminosity or size
(Kodama et al. 1998),
the similarity in the {\it C}-$\sigma$ relation we see above suggests that
Gonz\'alez sample would also share the same metallicity sequence.
This reinforces our coming assumption that the colour of the underlying
old population in ellipticals should locate along the {\it C}-$\sigma$
relation thus defined (see \S~3.2).
Although Larson, Tinsley, \& Caldwell (1980) claimed that the {\it C-M}
relation of field ellipticals are less tight than that of cluster ellipticals,
we do not find this in the case for the Gonz\'alez galaxies.
It is clearly needed to settle this problem with new systematic photometry
data of nearby field ellipticals.

The other notable thing is that the H$\beta$ strong galaxies are not always 
bluer in colours compared to the {\it C}-$\sigma$ relation, 
although they are expected to be appreciably bluer if suffering from 
the secondary burst of star formation recently. 
The 5 galaxies out of 10 H$\beta$ strong galaxies,
NGC 507, NGC 584, NGC 1700, NGC 3377, and NGC 7454,
locate below (bluer) the line of {\it C}-$\sigma$ relation in
Fig.~\ref{fig:uv}.
These galaxies are likely to have considerable number of young stars.
However, except NGC 3377, these galaxies are bluer no more than 0.1~mag,
which indicates little contamination of young stars to the galactic light.
The H$\beta$ strong galaxies should contain certain amount of young stars
and yet some of them do not show any evidence in colours for young stars.
In the following \S 3, we build a series of models for elliptical galaxies
with the young population superposed onto the underlying old
populations and examine how much young stars can be in hiding to give
compatible H$\beta$ line strengths while keeping the {\it C}-$\sigma$
relation tight.

\section{Model}

\subsection {Elliptical galaxy models}

We build models for elliptical galaxies, having experienced 
the secondary burst of star formation, 
by superposing the young stellar population onto the old underlying
populations. For the underlying populations, we adopt the
metallicity sequence models of elliptical galaxies 
calculated by Kodama \& Arimoto
(1997) and Kodama (1997). These models are calibrated to reproduce the
{\it C-M} relation of elliptical galaxies in Coma cluster by changing
a galactic wind epoch as a function of initial mass of a galaxy.
This is equivalent to change the mean stellar metallicity of galaxies along 
the {\it C-M} relation. The age of the galaxies is assumed to be 12~Gyr.
All stars formed very quickly before the wind ($<$0.5~Gyr).
The time scale of star formation is assumed to be 0.1~Gyr. 
A Salpeter-like initial mass function (IMF) is assumed with a slope 
$x=1.10$ (the Salpeter IMF corresponds to $x=1.35$ in our definition). 
The lower and the upper mass 
cutoffs $m_{\ell}$ and $m_{u}$ are set to be 0.1 and 60 $M_{\odot}$,
respectively. The mean stellar metallicity $\langle\log Z/Z_{\odot}\rangle$
decreases from 0.0 to $-$0.5 for the range of 6 magnitudes covered by the
{\it C-M} relation ($M_V=-23.0$ to $-17.0$~mag).
Chemical evolution is taken into account under the 
so-called infall model scheme (\cf K\"oppen \& Arimoto 1990).
The models are shown to match very well with the observed {\it C-M} relations
of elliptical galaxies in distant clusters
(Kodama \& Arimoto 1997; Kodama et al. 1998).

The velocity dispersion is assigned to each model galaxy
with a help of the $(U-V)_{e,0}$ versus $\log
\sigma_0$ relation defined in \S 2.3 (solid line in Fig.~\ref{fig:uv}) as:

\begin{equation}
\log \sigma_{0} = 2.024 [(U-V)_{e,0}-0.339],
\end{equation}
where we use theoretical $U-V$ for $(U-V)_{e,0}$.
The Lick spectral indices are calculated by using the polynomial 
analytical fits given by Worthey, Faber \& Gonz\'alez (1992).
The synthesized H$\beta$ indices of the underlying galaxy 
models with various initial masses are shown in Fig.~\ref{fig:hbeta} by the
solid line.

\subsection {Secondary star burst models}

\begin{figure}
\begin{center}
  \leavevmode
  \epsfxsize 1.0\hsize
  \epsffile{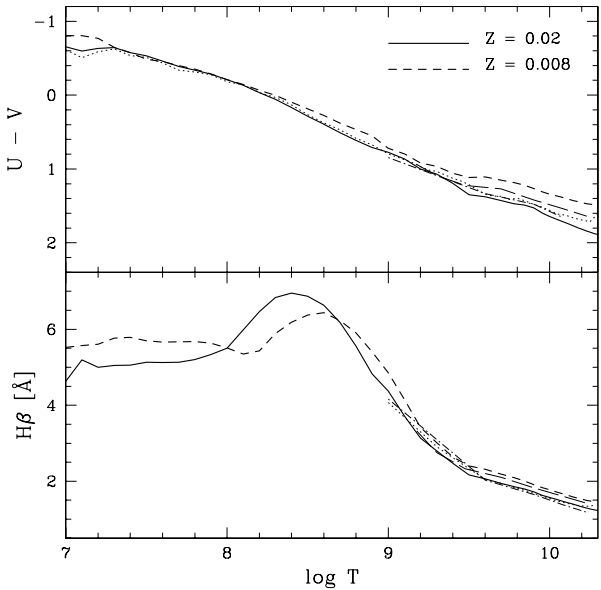}
\end{center}
\caption
{The $U-V$ and H$\beta$ evolution for Simple Stellar Population (SSP) models
with $Z=$0.008 (thick dashed lines) and $Z=0.02$ (thick isolid lines).
The other models with solar abundance ($Z=0.02$) from various authors
are also plotted for comparison. Light dotted lines, light
short-dashed lines, and long dash-dotted lines
correspond to SSP models of Bressan, Chiosi \& Tantalo (1996),
Vazdekis \etal (1996), and Worthey (1994), respectively.
}
\label{fig:uv_hbeta}
\end{figure}

We assume that the secondary burst happened
instantaneously. In such a case, the spectrum of the burst population is 
approximated by the so-called simple stellar 
population (SSP) model (\cf Buzzoni 1989). Thus we ignore the metal
enrichment during the secondary burst. The IMF slope of the secondary
population is assumed to be $x=1.35$ 
(the Salpeter IMF). $m_{\ell}$ and $m_{u}$ 
are the same as those of the underlying population.
Note we adopt different IMF slope for the burst popualtion from that of
the underlying population, simply because there is no reliable estimate
available.
However, our conclusion would not be affected
by this choice of IMF slope. If we were to use $x=1.10$ for the burst
population instead, the acceptable burst strength would become even smaller
due to more numerous young stars. Therefore, our coming conclusion would be
more strengthened.
The evolution of $U-V$ and
H$\beta$ index of the SSP models are illustrated in Fig.~\ref{fig:uv_hbeta}. 
The thick dashed lines and the thick solid ones represent the SSP models
with $Z=0.008$ and $0.02$, respectively. 
The thin lines represent various SSP models with the solar abundance 
($Z \simeq$ 0.02)
available in the literature (Worthey 1994; Bressan et al. 1996; Vazdekis et 
al. 1997). The age of the burst population $T_{b}$ is varied from 0.1 
Gyr to 5~Gyr as a free parameter.
If the secondary burst is induced by capturing a small
gas-rich galaxy, as is most likely, the mean stellar metallicity of the
burst population should not exceed the solar metallicity,
since typical metallicities of dwarf irregulars
are well below the solar value (Skillman, Kennicutt \& Hodge 1989).
Even if the chemical enhancement during the burst is considered, the
average metallicity of the bust population 
hardly exceeds the solar value unless the IMF slope is significantly flatter
than the Salpeter IMF.
Therefore we adopt $Z_{b}=0.008$ and $Z_{b}=0.02$ as the
representative metallicities of the secondary burst.
The burst strength ($f_{b}$) is defined by a {\it mass} fraction of burst
population to the whole galaxy at the present epoch (12~Gyr), 
and are allowed to vary from 0.1~\% to 20~\%.
We assume that the velocity dispersion does not change before and after the
burst, since the mass involved in the secondary burst is
usually small compared to that of the
underlying population. Even if the fraction of the burst is as high as 
20~\% in mass, the resulting increase of the velocity dispersion is 
at most $\Delta \log \sigma_0 =0.05$ if the galaxy is in virial 
equilibrium before and after the burst.
If the dark
matter dominate the galaxy potential, the influence of the burst population
to the velocity dispersion should be less significant.

Along the {\it C}-$\sigma$ relation,
the young SSP model is superposed in such a way that
the mass fraction of young population
to the total mass equals to a given burst fraction $f_{b}$.
The $U-V$ colour and H$\beta$ index of a total galaxy are calculated
from the resulting composite spectrum.

\section{Comparison}

\begin{figure}
\begin{center}
  \leavevmode
  \epsfxsize 1.0\hsize
  \epsffile{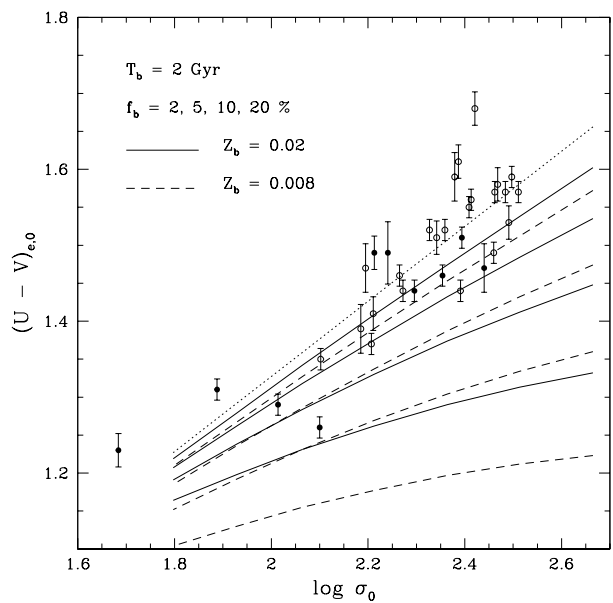}
\end{center}
\caption
{Models in the $(U-V)_{e,0}$ versus $\log \sigma_{0}$ for elliptical galaxies
 with the secondary star burst. Circles show our sampled galaxies and the
 dotted line gives the regression line for galaxies shown in 
 Fig.2. The dashed line and the solid one indicate models 
 with the secondary star burst of $Z_{b}=0.008$ and $Z_{b}=0.02$,
 respectively. Burst age $T_{b}$ is assumed to be 2.0~Gyr. The burst
 strength $f_{b}$ is increased from 2~\% (top) to 20~\% (bottom).}
\label{fig:uv_tb2}
\end{figure}

\begin{figure}
\begin{center}
  \leavevmode
  \epsfxsize 1.0\hsize
  \epsffile{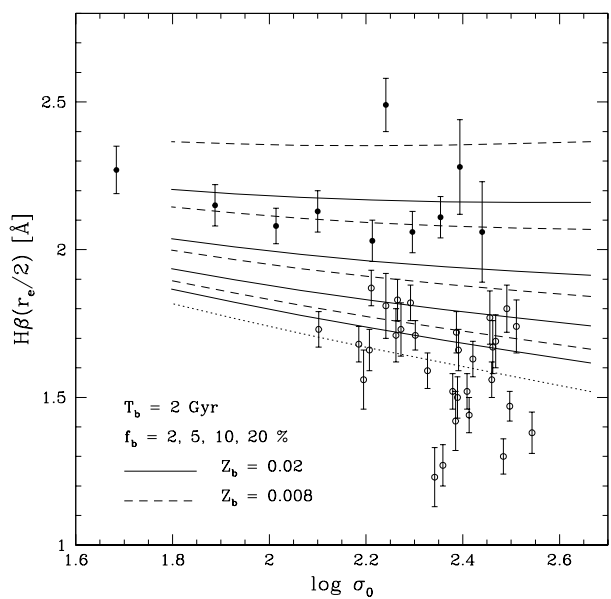}
\end{center}
\caption
{The same as Fig.~\ref{fig:uv_tb2},
but H$\beta$ versus $\log \sigma_{0}$ diagram.
The dotted line shows the sequence of underlying galaxy models, the same as
the solid line in Fig.~\ref{fig:hbeta}.
The burst strength $f_{b}$ is increased from bottom to top direction.}
\label{fig:hbeta_tb2}
\end{figure}

Figure~\ref{fig:uv_tb2} illustrates the colour change due to an increase of
the burst strength in the $(U-V)_{e,0}$ versus $\log \sigma_{0}$ diagram.
Burst age $T_{b}$ is fixed to be 2~Gyr.
The open and filled circles show the same galaxies as in Fig.~\ref{fig:uv}, 
and the dotted line gives the regression line for the data, along which we
assign the underlying galaxies.
The dashed line and the solid one correspond to 
the metallicity of the burst $Z_{b}=$0.008 and
0.02, respectively.
The burst strength $f_{b}$ is increasing from the top to the bottom.
Apart from one exception, NGC 3377, the maximum burst strength allowed
to be consistent with the data is less than $\sim$ 7~\% for $T_{b}=$2~Gyr.
In the same way, models with various combination of $T_{b}$, $f_{b}$, and
$Z_{b}$ are confronted to the data and the upper limit of the 
burst strength 
is investigated for each case. The results are summarized in Table 1.
The acceptable amount of colour change $\Delta (U-V)$
due to the young population is set to $-$0.1 at log $\sigma_0=2.35$.
The maximum burst strength $f_{b}$ is adjusted in 
such a way that $\Delta (U-V)$ reaches this limit for a given set of 
$T_{b}$ and $Z_{b}$.
As a result, we find that the permitted range of the burst strength
should be very small; typically less than a few percent in mass,
at most 4--7~\% within recent 2~Gyr.

 \begin{table*}
   \begin{center}
   \caption{The constraint on the recent secondary burst of star formation
            in elliptical galaxies.
            The upper limit on $f_b$ given in column (3)
            is set in such a way that
            the colour change of galaxies in $U-V$
            due to the superposition of young population (column 4) is $-$0.1
            at log $\sigma_0$=2.35.
            Columns (1) and (2) give metallicity and age of the burst
            population, respectively.
            The corresponding H$\beta$ indices and the
            luminosity weighted mean stellar ages of galaxies are given
            in columns (5)-(7). See text for details.}
   \vspace{0.4cm}
   \label{table-1}
   \begin{tabular}{l|ll|cccc}
   \hline
     $Z_{b}$ & $T_{b}$ & $f_{b}$ & $\Delta (U-V)$ & H$\beta$ &
     $\langle T \rangle_{U}$ & $\langle T \rangle_{V}$\\
             & (Gyr)  & (\%)  & (mag)          & (\AA)    & (Gyr)    & (Gyr) \\
   \hline\hline
           & 0.2 & 0.18 & $-$0.10 & 1.84 & 7.3 & 10.4\\
           & 0.5 & 0.6  & $-$0.10 & 1.94 & 7.5 & 9.8\\
     0.02  & 1   & 1.9  & $-$0.10 & 1.92 & 7.6 & 9.3\\
           & 2   & 7    & $-$0.10 & 1.86 & 7.4 & 8.4\\
           & 5   & $>$20& ---     & ---  & --- & --- \\
   \hline
           & 0.2 & 0.15 & $-$0.10 & 1.79 & 7.4 & 10.5\\
           & 0.5 & 0.45 & $-$0.10 & 1.90 & 7.7 & 10.1\\
     0.008 & 1   & 1.3  & $-$0.10 & 1.94 & 7.8 & 9.6\\
           & 2   & 4    & $-$0.10 & 1.84 & 8.0 & 9.2\\
           & 5   &14    & $-$0.10 & 1.76 & 9.1 & 9.7\\
   \hline
   \end{tabular}
   \end{center}
 \end{table*}
 
The above number is considerably smaller than that 
given by de Jong \& Davies (1996), who obtained 10--15~\% 
young population for $T_{b}=2$~Gyr burst to
reproduce the strong H$\beta$ line strengths ($>2$~\AA).
Independently, as described in \S~3, we calculated H$\beta$ indices
for the secondary star burst models to see the consistency.
The results are shown in Fig.~\ref{fig:hbeta_tb2} for the case of
$T_b=$2~Gyr.
It is clearly seen that the H$\beta$ strong galaxies can only be reproduced
by adding more than 10--15~\% young star burst, which is consistent with
de Jong \& Davies (1996).
However, this number is more than twice as much as that expected
from the $U-V$ analysis presented above.
We also estimated H$\beta$ values that correspond to $f_{b}$ 
values constrained from $\Delta (U-V)$.
The results are tabulated in the fifth column of Table 1.
As a consequence, the H$\beta$ indices are always smaller than 2~\AA.
Thus the models constrained from the $U-V$ colours fail to reproduce
the H$\beta$ strong galaxies (H$\beta$ $>$ 2~\AA).
Since the strong Balmer absorption lines are 
characteristic of A-type stars, the contribution from such stars becomes
maximum at around $\sim$ 0.2--0.5~Gyr after the onset of burst
(Fig.~\ref{fig:uv_hbeta}).
However, even with $T_b \simeq$ 0.2--0.5~Gyr, the resulting
H$\beta$ indices are again less than 2~\AA.
True one can obtain stronger H$\beta$ strengths by imposing 
much larger burst strength $f_b$, but the resulting galaxy becomes 
too blue in $U-V$. We will discuss this $U-V$ versus H$\beta$ discrepancy
later.

If the age of secondary burst is as young as 0.5--1~Gyr
as suggested from the numerical simulation 
of shells around dynamically disturbed
ellipticals (Hernquist \& Quinn 1987),
the burst strength should be quite small, such as less than 1--2~\%.
Conversely, if we assume older burst ages, the limit of $f_{b}$
becomes larger (more than 20~\% for $T_{b}=$5~Gyr, for example),
but $(U-V)$ - H$\beta$ discrepancy becomes larger.

For reference,
the $U$- and $V$-band luminosity weighted mean age of stars in
a galaxy $\langle T \rangle_{U,V}$ are also given in Table 1,
which are defined as:
 \begin{equation}
 \log \langle T \rangle_{U,V} = \frac{\Sigma (\log T) L_{U,V}}{\Sigma L_{U,V}},
 \end{equation}
where summation is taken for all stars in a galaxy.
Thus, the lower limit of $\langle T \rangle_{U}$ is 7.3--9.1~Gyr and that of
$\langle T \rangle_{V}$ is 8.4--10.5~Gyr.
These are much older than the age of bulk of ellipticals that
Worthey, Trager \& Faber (1996) assigned from the Balmer line indices.
These authors find significant number of extremely young elliptical
galaxies with age only 2--3~Gyr.
The mean age is also quite young such as 4--5~Gyr, nearly half of what we 
find.
The discrepancy of the results between the analysis based on the colours
and the one based on the line indices can be seen here as well.

In conclusion, apart from NGC 3377, which could be the only
candidate that might actually have high burst strength of upto 15--20~
\% for $T_{b}=2$~Gyr and could be consistent with its H$\beta$ strength,
the intensity of the secondary star burst should be less
than only a few percent in mass to be compatible with the observed 
$U-V$. It is not strong enough to explain the 
H$\beta$ strong galaxies, however.
 
\section{Discussion and conclusions}

So far we have calibrated the models of the underlying old populations
using the linear regression line of the data on the {\it C}-$\sigma$
diagram, supported by the existence of mass-metallicity relation
for elliptical galaxies as we discussed in \S~2.3.
However, one could alternatively take the location of the underlying
population as the reddest envelope of the sample galaxies,
if all galaxies bluer than
the red envelope are regarded as being contaminated by young stars.
Although we believe this is unlikely as discussed above, we estimate
how much the upper limit of $f_b$ could be increased if this effect
is the case.
If we take the red envelope at 0.07~mag redder than the
regression line, the acceptable burst strength increases to $f_b\sim$10~\%.
Therefore, even in this case, the maximum burst strength in mass fraction
should be less than 10~\% at most, within recent 2~Gyr.

\begin{figure}
\begin{center}
  \leavevmode
  \epsfxsize 1.0\hsize
  \epsffile{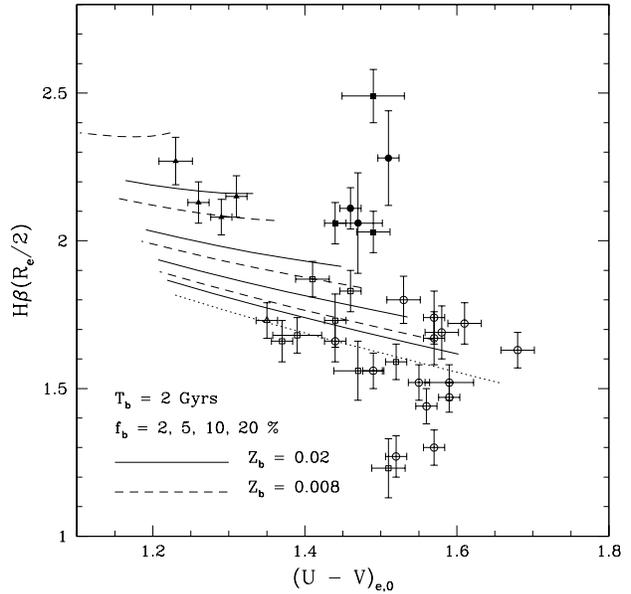}
\end{center}
\caption
{The same as Fig.~\ref{fig:uv_tb2},
but H$\beta$ versus $(U-V)_{e,0}$ diagram.
The range of $\sigma_0$ is discriminated by the type of symbols:
Triangles, squares, and circles correspond to $\log \sigma_0<2.15$,
$2.15<\log \sigma_0<2.35$, and $\log \sigma_0>2.35$, respectively.
The dotted line shows the sequence of underlying galaxy models.
The burst strength $f_{b}$ is increased from bottom to top direction.}
\label{fig:uv_hbeta_tb2}
\end{figure}

Furthermore, one may claim that low mass systems have more often disky
shapes and can contain more young stellar populations
(e.g., Kormendy and Bender 1996), therefore the slope of the
{\it C}-$\sigma$ relation of the underlying old population might possibly
be flattened at the faint end than that defined in this paper.
This is equivalent to
weaken the metallicity variation as a function of galaxy mass and to
introduce some age difference instead.
In fact, as shown in Fig.~\ref{fig:uv_hbeta_tb2} where
the sample galaxies are plotted in the $U-V$ versus H$\beta$ diagram
with the secondary burst models superposed,
the four H$\beta$ strong ellipticals with low dynamical masses
($\log \sigma_0 < 2.15$) could be explained consistently both in
$U-V$ and H$\beta$ at the same time with some combination of metallicity
and burst strength, if one could neglect the {\it C}-$\sigma$ relation.
The other H$\beta$ strong ellipticals with larger masses
locate out of the model grid, and the $(U-V)$-H$\beta$ discrepancy is
quite evident for these galaxies in anycase.
However, if lower mass systems tend to be more contaminated
by young stars, the slope of the {\it C}-$\sigma$ relation and hence
the {\it C-M} relation is expected to vary from cluster to cluster,
given the colour is very sensitive to a small difference of the amount
of young stars. This trend is far from what we actually observe both locally
and at high redshifts
(Garilli et al. 1996, Stanford et al. 1997, Kodama et al. 1998,
L\'opez-Cruz 1997). It has been shown that most of clusters have
a common slope when compared to Coma or Virgo cluster.
Even if it were the case that low mass galaxies would contain more young stars,
the $(U-V)$-H$\beta$ discrepancy could not be compensated.
In order to be consistent with strong H$\beta$ indices of these low mass
galaxies, the secondary star burst of 15--20~\% in mass is required
with $T_b=2$~Gyr.
If they were to have such strong star burst (15~\%), the colour of the
underlying population should be $U-V\simeq1.5$, just comparable to the typical
colour of the rest of ellipticals, and we would no longer see the
{\it C}-$\sigma$ relation for the underlying old population.
The 15~\% burst with $T_b=2$~Gyr gives luminosity weighted ages of
$\langle T \rangle_U=5.2$ and $\langle T \rangle_V=6.2$, which are 6--7~Gyr
younger than the bulk of more massive ellipticals
(which corresponds to about 4 magnitude brighter using the Fabor-Jackson law).
As expected, the 6--7~Gyr difference in 4 magnitude range of {\it C-M} relation
is almost comparable to a pure age sequence discussed in Kodama \& Arimoto
(1997) and Kodama et al. (1998),
which is absolutely rejected by the evolution of {\it C-M} slope
as a function of redshift.
Therefore, the observed {\it C-M} relation of clusters does not support the
idea xthat the low mass H$\beta$ strong
ellipticals have experienced much stronger secondary burst than our estimate.

After all, the $(U-V)$-H$\beta$ discrepancy does remain; i.e.,
the integrated $U-V$ colour of Gonzalez's sample of ellipticals
permit only 10\% of the secondary stellar population at most,
while the strong H$\beta$ indices suggest much larger burst strengths.
We here consider possible reasons for this discrepancy.

(1) {\it Model uncertainty}. One possible reason is
that the H$\beta$ synthesis model based on the empirical
calibration of Worthey \etal (1992) may be still premature partly because
the scatter around the polynomial fitting lines might be too large
and/or because the zero-point of the model might be uncertain.
$U-V$ could have uncertainties as well (Charlot, Worthey \&
Bressan 1996),
but they are expected to be small at ages 1--2~Gyr or younger,
since at such young ages, $U-V$ colour is driven primarily by the evolution
of the main sequence turn-off stars, and is relatively simple to model.

(2) {\it Emission correction}.
Gonz\'alez (1993) corrected the contamination of H$\beta$ emission for
the equivalent width of H$\beta$ absorption by using the intensity of
[OII] emission. Although this empirical procedure is reasonable, it
causes uncertainty in the H$\beta$ indices.
If the correction tends to be overestimated, the real H$\beta$ absorption
could be much lower and consistent with $U-V$ analysis.

(3) {\it Aperture effect}.
H$\beta$ indices are defined in the $r_e/2$ aperture, while the colours
are defined in the $r_e$ aperture.
If all stars of the secondary population are formed locally within $r_e/2$,
the allowed burst strength 
could be 1.56 times larger within $r_e/2$ following de Vaucouleurs (1948)
$r^{1/4}$-law. This could rise the model H$\beta$ indices by about 0.1--0.15~
\AA, and it could marginally make up the discrepancy.
According to Davies, Sadler \& Peletier (1993), however, 
H$\beta$ absorption is constant or increasing with a radius in almost
all cases of elliptical galaxies in their sample.
Though this could be partly due to the dilution of the H$\beta$ absorption
feature by emission in the centers of galaxies (Davies et al. 1993), 
H$\beta$ would be hardly changed between $r_e/2$ and $r_e$ aperture.

\begin{figure}
\begin{center}
  \leavevmode
  \epsfxsize 1.0\hsize
  \epsffile{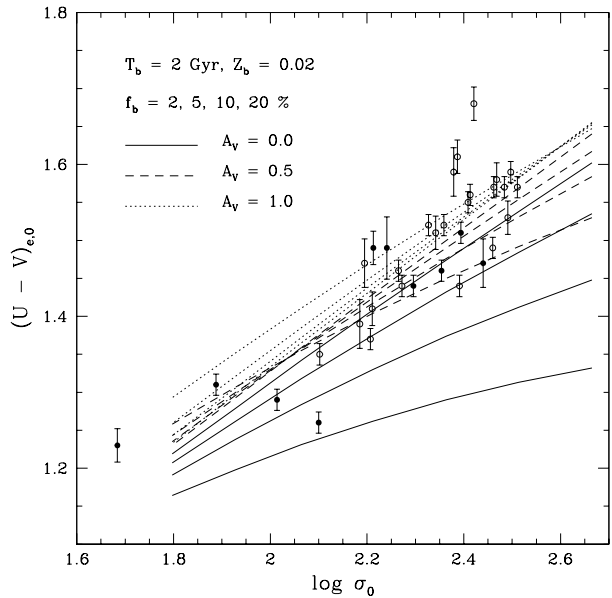}
\end{center}
\caption
{
 The effect of interstellar extinction on $(U-V)_{e,0}$ 
 versus $\log \sigma_{0}$
 diagram.
 The solid line corresponds to the reddening free burst models with
 $T_{b}=2$Gyr and $Z_{b}=$0.02, the same as Fig~\ref{fig:uv_tb2}.
 The dashed and dotted lines indicate the models with $A_V=$0.5~mag and
 1~mag, respectively.
}
\label{fig:uv_av_tb2}
\end{figure}

\begin{figure}.
\begin{center}
  \leavevmode
  \epsfxsize 1.0\hsize
  \epsffile{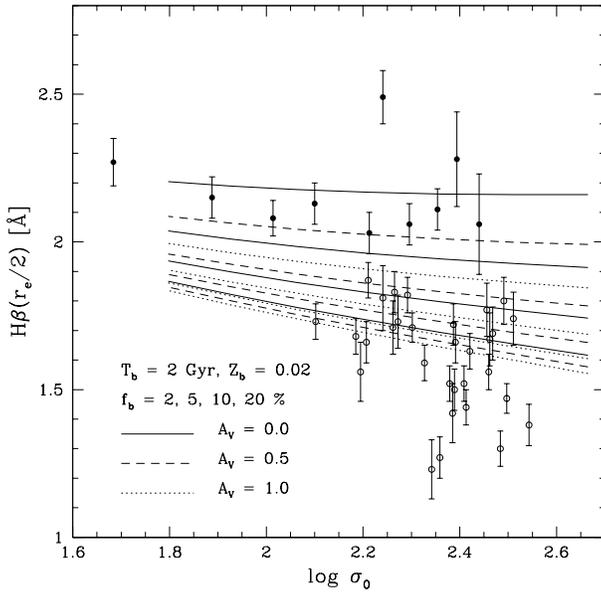}
\end{center}
\caption
{
The same as Fig.~\ref{fig:uv_av_tb2},
but H$\beta$ versus $\log \sigma_{0}$ diagram
The burst strength $f_{b}$ is increased from bottom to top direction.}

\label{fig:hbeta_av_tb2}
\end{figure}

(4) {\it Internal dust extinction}.
The internal dust extinction is expected to have more effect on
$U-V$ colour than on H$\beta$ indices.
It would redden $U-V$ of galaxies without reducing H$\beta$ indices
much and could dissolve the discrepancy.
To see the effect in some more detail, the internal extinction is taken into
account to the model spectra in a simple way. 
The extinction law is taken from Mathis (1990).
We assumed that the dust distribution in the galaxies is localized only
in the region where the secondary burst occurred.
Hence we applied the extinction correction only for the burst population,
and not for the old underlying population.
In Figures~\ref{fig:uv_av_tb2} and \ref{fig:hbeta_av_tb2},
the effect of extinction is shown for $U-V$ and H$\beta$,
respectively. $A_V$ is changed as 0.0, 0.5, and 1.0 mag.
The model with $f_b$=20~\% and $A_V$=0.5~mag, for example,
could possibly reproduce some of the H$\beta$ strong galaxies both in
$U-V$ and H$\beta$.
The effect of internal extinction certainly allows a higher upper limit to
the burst strength.
A study of infrared colours of Gonz\'alez
sample would help since the SED at longer wavelengths is less vulnerable
to the dust extinction. 

In summary, although considerable dust extinction for burst population could
solve the discrepancy between $U-V$ and H$\beta$ analyses, we rely on $U-V$
at the moment, considering that the H$\beta$ might have great uncertainties
both in models and in observational data.
In this case, we can put strong constraint on the secondary episode
of star formation in ellipticals:
Elliptical galaxies, regardless of their environments, 
experience little secondary
star formation in the recent past. For example, the mass fraction of the
young burst population should be quite small, typically less than 4--7~\%,
and 10~\% at most within recent 2~Gyr,
This suggests that the dynamical disturbance found in some ellipticals
is more likely to be caused by a capture of dwarf galaxy
with a mass of order $10^{9}$--$10^{10} M_{\odot}$, rather than a merging of
two galaxies of comparable size.
If ellipticals formed by the recent mergings of
comparable galaxies, as hierarchical clustering models 
of the universe suggest,
it should have proceeded without any significant star formation.
However, it is yet to be studied in detail how such process can create the observed
{\it C-M} relation of ellipticals.
 
\section*{Acknowledgements}

We would like to thank 
A. Arag\'on-Salamanca, R.L. Davies, R.G. Bower, R.S. de Jong, 
H. Kuntschner, T.J. Ponman, and D.A. Forbes for very useful discussions.
T.K. thanks the Japan Society for Promotion of Science
(JSPS) Postdoctoral Fellowships for Research Abroad for supporting his stay
in the Institute of Astronomy, Cambridge, UK. N.A. thanks the JSPS for
supporting his stay in Observatoire de Paris, 
Section de Meudon, France.
This work was financially supported in part by a Grant-in-Aid for the
Scientific Research (No.09640311) by the Japanese Ministry of Education,
Culture, Sports and Science.

\end{document}